\documentclass[a4paper,12pt]{article}
\usepackage{graphicx}
\usepackage[T1,T2A]{fontenc}
\usepackage[utf8]{inputenc}
\usepackage{amssymb}
\usepackage{amsmath}
\usepackage{setspace}
\usepackage{xcolor}
\graphicspath{{images/}}
\usepackage[font=normalsize,labelfont=sf,textfont=sf]{subfig}

\begin{document}

\thispagestyle{empty}

\begin{center}
{\large
{\bf 
Critical properties of a 2D frustrated magnet with non-magnetic impurities
} 
}
\vskip0.5\baselineskip{
\bf 
D.~N.~Yasinskaya$^{*1}$, 
V.~A.~Ulitko$^{1}$,
Yu.~D.~Panov$^{1}$
}
\vskip0.1\baselineskip{
$^{1}$Ural Federal University 620002, 19 Mira Street,  Ekaterinburg, Russia
}
\vskip0.1\baselineskip{
$^{*}$daria.iasinskaia@urfu.ru
}
\end{center}

We report on classical Monte Carlo study of phase transitions and critical behavior of a 2D spin-pseudospin model describing a
dilute magnet with competing charge and spin interactions. The static critical exponents of the specific heat and correlation length
are calculated using the finite-size scaling theory in a wide range of model parameters. The order of phase transitions is analyzed
within the energy histogram method. It is found that approaching the frustration point and increasing the density of non-magnetic
impurities leads to non-universal critical behavior and first-order phase transitions. Features of non-universal critical behavior are
shown to depend on the relationship between the parameters of the spin and pseudospin interactions.\\


\section{Introduction}
The study of phase transitions (PT) and critical phenomena in spin systems is of great importance for modern condensed matter physics. In particular, frustrated and diluted magnetic systems are a subject of increased attention. Frustration causes a
variety of unstable phase states, significant and highly unpredictable changes in the critical, thermodynamic and magnetic properties~\cite{Diep,Zarubin,Kalz}. In addition, magnetic systems with frustrations cover a wide range of objects with special magnetic states, such as spin liquid, spin glass~\cite{Balents2010} and spin ice~\cite{Bramwell2001}. The study of dilute magnets is significant not only
from a theoretical, but also from a practical point of view,
since it expands the understanding of behavior and properties
of real materials, which, as a rule, have defects and impurities.
Moreover, disorder caused by doping impurities and defects
significantly affects the critical behavior and phase states of
magnetic systems~\cite{Novotny,Dotsenko,Murtazaev}.

The focus of theoretical research aims to study more realistic complex models. The study of phase transitions and critical
phenomena in such systems by methods of theoretical physics
is a rather difficult problem. Therefore, complex systems are
intensively investigated by numerical-simulation methods such
as Monte Carlo methods. For example, Monte Carlo methods
have been used to study magnetic multilayers~\cite{Masrour2016, Masrour2018, Belhamra2020}, Kekulene structures~\cite{JabarKekulene1,JabarKekulene2}, decorated lattices~\cite{MasrourDecor1,MasrourDecor2} and many other various systems.

One of the most effective methods for studying complex
magnetic systems with different degrees of freedom is the
pseudospin formalism. Pseudospin models are widely used
for describing the properties of binary alloys, classical and
quantum liquids, dilute magnets, superconductors, and many
other physical systems~\cite{Sivadiere,Cannas}.  Here in the paper, we
consider the spin-pseudospin model to explore the site-dilute
2D magnet with frustration. The frustration is caused by
competitive charge (pseudospin-1) and magnetic (spin-1/2)
orders. On the one hand, this model generalizes the frustrated
mixed-spin Ising model with nonmagnetic impurities. On
the other hand, this model was initially proposed~\cite{SPSModel} to consider this competition of different orders in underdoped high-$T_c$ cuprates like La$_{2-x}$Sr$_x$CuO$_4$ in the normal state. The
properties of the spin-pseudospin model in the ground state
were studied within the mean-field approximation~\cite{SPSModel}, the
temperature phase diagrams were obtained within the Bethe
approximation~\cite{Bethe}.  It was shown that the ground state phase
diagrams and thermodynamic properties of the system are
different for the cases of weak and strong spin exchange
interactions. Complex effect of non-magnetic impurities and
frustration on features of the phase states formation in the
weak exchange limit was studied in the work~\cite{YasinskayaPhase}. The critical
behavior of the model was considered only in the strong spin
exchange limit~\cite{YasinskayaCrit}. In this work, we present the Monte Carlo
study of critical properties of an anisotropic dilute magnet
with competition between charge and magnetic orders in a
wide range of model parameters.

\section{Model and methods}

In the framework of the spin-pseudospin model~\cite{Moskvin} we consider the CuO$_2$ planes as a charge triplets system consisting of three many-electron mixed-valence centers [CuO$_4$]$^{7-,6-,5-}$ (nominally corresponding to the copper ion states Cu$^{1+,2+,3+}$). Two non-magnetic states Cu$^{1+,3+}$ are associated with two pseudospin projections $S_z=\pm 1$. The Cu$^{2+}$ state is associated with the $S_z=0$ projection of pseudospin and has a conventional spin $s=1/2$.
The Hamiltonian includes on-site density-density correlations in the form of single-ion pseudospin anisotropy ($\Delta$), inter-site density-density correlations in the form of pseudospin exchange interaction of Ising type ($V$), and Ising spin $s=1/2$ exchange coupling ($J$):
\begin{equation}
\mathcal {H} = 
\Delta \sum_i^{\phantom{N}} S_{iz}^2 
+ V \sum_{\left\langle ij\right\rangle} S_{iz} S_{jz} 
+ \tilde{J} \sum_{\left\langle ij\right\rangle}  \sigma_{iz} \sigma_{jz} - \mu \sum_i S_{iz},
\label{H}
\end{equation}
where both $V,J>0$. Here the sum runs over $N = L \times L$ sites of the square lattice, $\langle ij \rangle$ denotes the nearest neighbors, $\tilde{J}=Js^2$, $\sigma_{iz} = P_{0i} s_{iz}/s$ is a normalized $z$-component of the spin $s = 1/2$, multiplied by the projection operator $P_{0i} = 1-S^2_{iz}$, which distinguishes the magnetic Cu$^{2+}$-state with $S_z=0$.
$\mu$ is the chemical potential that is necessary to assume the constant charge constraint
\begin{equation}
n = \frac{1}{N} \sum\limits_i S_{iz} = const,
\label{const}
\end{equation}
where $n$ corresponds to the density of doped charge in the system. Thus, the system is diluted with nonmagnetic annealed interacting impurities acting as defects for the spin subsystem.

We used a parallel modification of the classical Monte Carlo (MC) method~\cite{Budrin}, which takes into account the condition~(\ref{const}). The temperature dependence of the specific heat is obtained using the following expression:

\begin{equation}
C = \frac{1}{N}\frac{\langle E^2 \rangle - \langle E \rangle^2}{T^2},
\end{equation}
where $E$ is the energy of the system with the Hamiltonian~(\ref{H}).
The order parameter $\mathcal{O}$ for checkerboard antiferromagnetic ($m$) and charge ordered ($M$) phases is determined as follows:
\begin{equation}
\mathcal{O} =
\left\{
\begin{array}{l}
m = m_1 - m_2,\\
M = M_1 - M_2.\\
\end{array}
\right.\\
\end{equation}
Here $m_{\lambda}=\sum\limits_{i \in \lambda} s_{iz}$ --  magnetization of the $\lambda$ sublattice, $M_{\lambda}=\sum\limits_{i \in \lambda} S_{iz}$ -- total charge of the $\lambda$ sublattice (pseudo-magnetization), $\lambda = 1,2$ denotes the checkerboard sublattice.

The Binder cumulant method~\cite{Binder1992} is used for accurate determination of PT temperatures. The temperature dependencies of cumulants
\begin{equation}
U_L=1-\frac{\langle \mathcal{O}^4 \rangle_L}{3 \langle \mathcal{O}^2 \rangle^2_L}
\end{equation}
for systems with different sizes $L$ intersect in the critical point $T_c$.

The critical exponents of the specific heat $\alpha$ and correlation length $\nu$ are calculated in the framework of the finite-size scaling theory. For sufficiently large $L$ one can write~\cite{Ferrenberg}
\begin{equation}
C (T=T_c,L) \sim L^{\alpha/\nu},
\label{Cscale}
\end{equation}
\begin{equation}
V_n (T=T_c,L) \sim L^{1/\nu},
\label{Vnscale}
\end{equation}
where $V_n$ is the logarithmic derivative of the order parameter $\mathcal{O}^n$:
\begin{equation}
V_n=\frac{\langle \mathcal{O}^n E \rangle}{\langle \mathcal{O}^n \rangle} - \langle E \rangle.
\end{equation}
For the 2D Ising model regime of the critical behavior, the scaling relation for the specific heat will be the logarithmic one:
\begin{equation}
C(T=T_c,L) \sim \ln L.
\label{log}
\end{equation}
In this case, we assume $\alpha=0$~\cite{Novotny}. Calculation of critical exponents was performed for the system with periodic boundary conditions, $L = 32$-$64$, and the total number of MC steps was $22 \cdot 10^6$. All calculations were performed simultaneously for $10^2$ copies of the system in order to improve the accuracy of the results.

To determine the order of PT, we performed a precise analysis of energy histograms~\cite{Ferrenberg}. Energy histograms were plotted for $L = 128$. Data binning was performed by dividing $2 \cdot 10^6$ MC steps into a thousand of intervals.

\section{Monte Carlo results}

\begin{figure}[h!]
\centering
\includegraphics[width=0.8\linewidth]{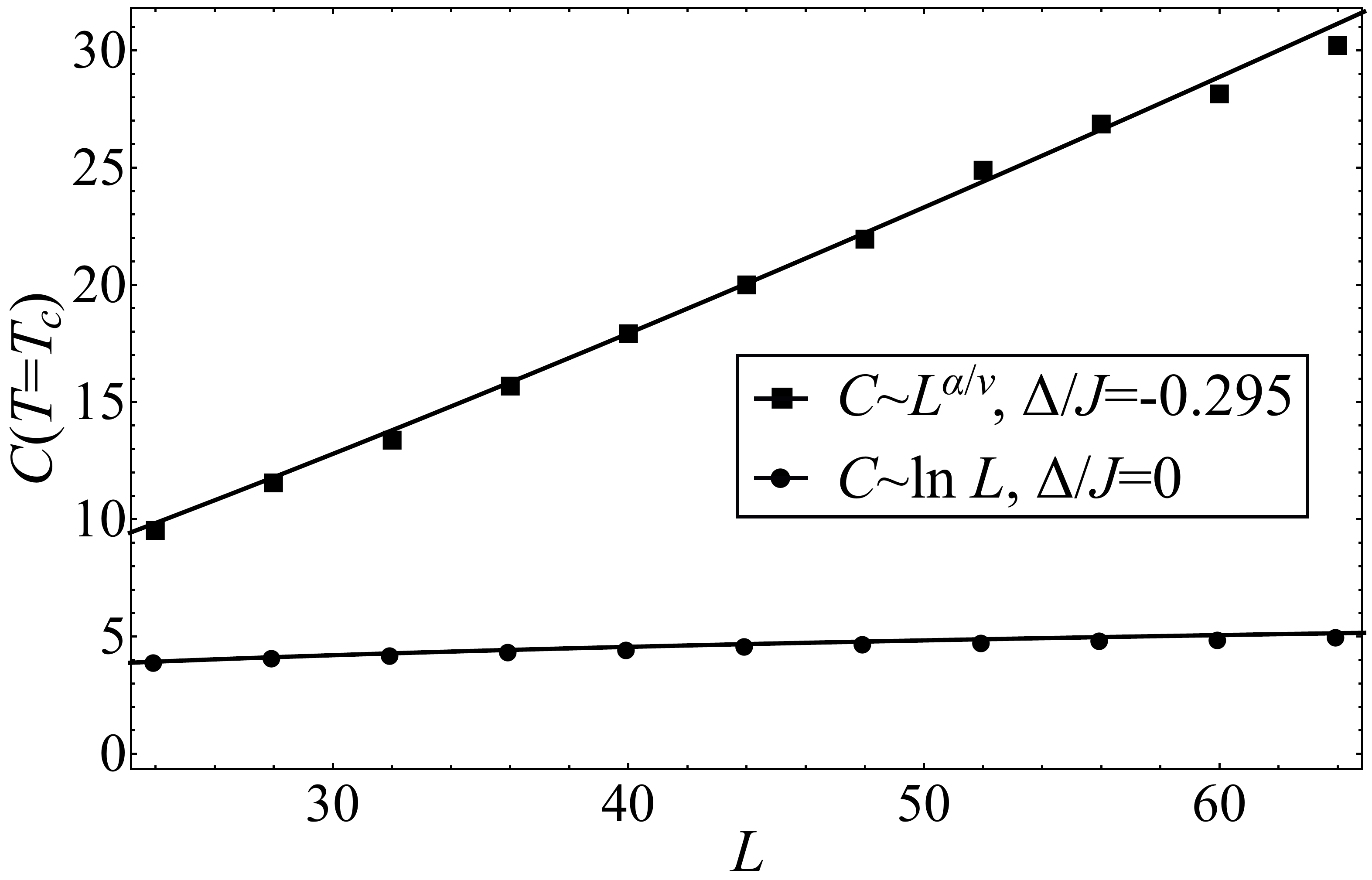}
\caption{Dependencies of specific heat $C$ at the critical point $T_c$ on lattice sizes $L$ for $V/\tilde{J}=0.4$, $n=0$. Logarithmic scaling relation for $\Delta/J = 0$ yields $\alpha = 0 \pm 0.004$, power relation for $\Delta/J = -0.295$ yields $\alpha/\nu = 1.174 \pm 0.018$}
\label{ascale}
\end{figure}

Fig.~\ref{ascale} shows the dependencies of specific heat $C$ on lattice linear sizes $L$ for $V/\tilde{J}=0.4$, $n=0$ and two different values of $\Delta/J$. For $\Delta/J=0$, the specific heat behaves according to the logarithmic law~(\ref{log}), which yields $\alpha = 0 \pm 0.004$. For $\Delta/J = -0.295$, the specific heat is scaled according to the power law~(\ref{Cscale}), which corresponds to $\alpha/\nu = 1.174 \pm 0.018$.

\begin{figure}[h!]
\centering
\includegraphics[width=0.8\linewidth]{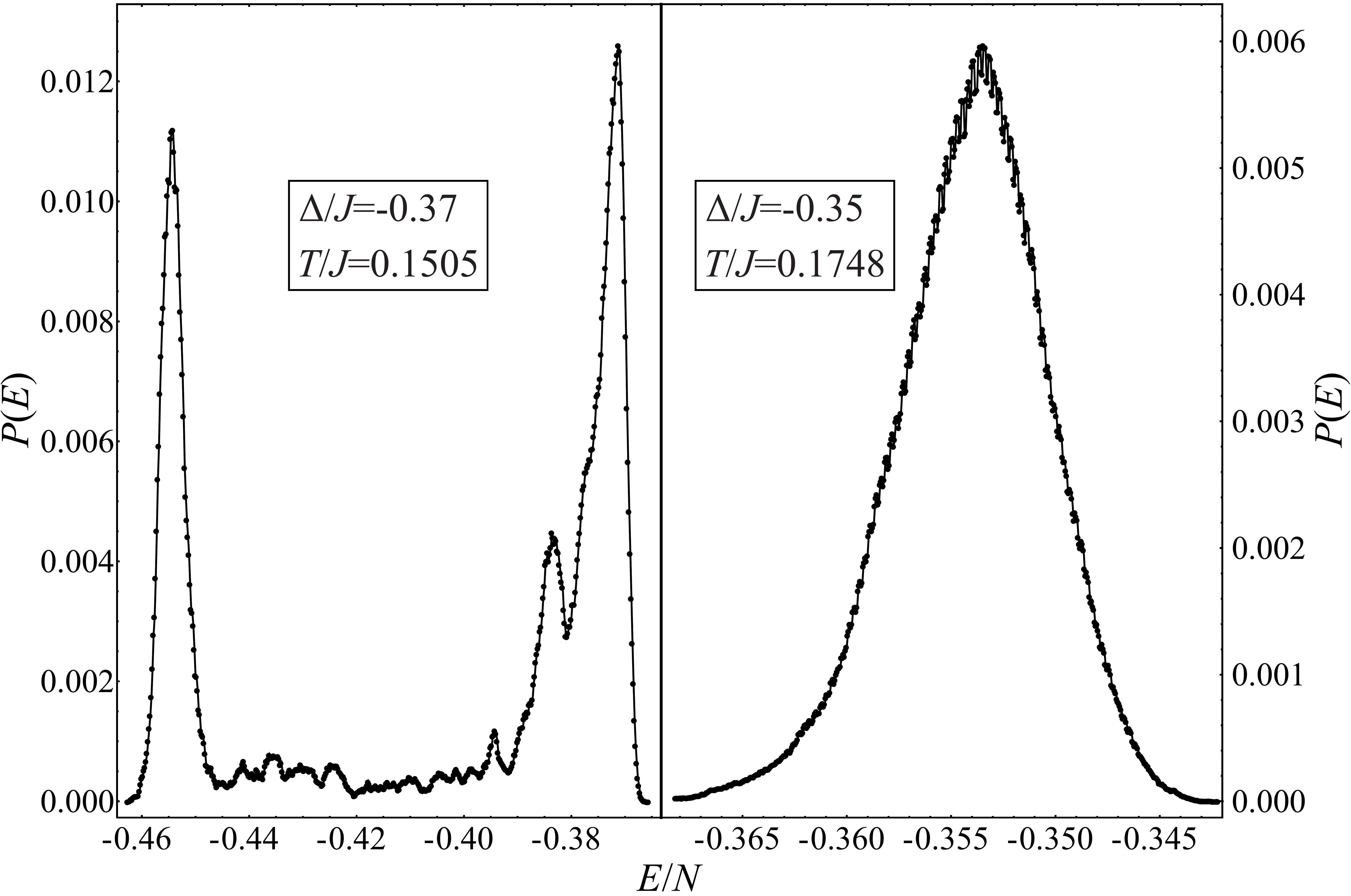}
\caption{Histograms for $V/\tilde{J} = 0.2$, $n = 0$, $L = 128$. For $\Delta/J = -0.37$ (left), the two-peak histogram provides a first-order phase transition. One-peak Gaussian histogram for $\Delta/ J = -0.35$ (right) indicates a second-order phase transition}
\label{histo}
\end{figure}

It is known that the order of PT can change in many frustrated systems~\cite{Diep, Kalz}. Fig.~\ref{histo} exhibits histograms of the energy distribution for $V/\tilde{J}=0.2$, $n=0$ and two different values of $\Delta/J$. The histograms are plotted near the critical temperature. For $\Delta/J = -0.37$ at $T = 0.1505$, the histogram has a double-peak feature, which indicates a first-order PT. For $\Delta / J = -0.35$ at $T = 0.1748$, the histogram has only one maximum. This was interpreted as evidence for a second-order PT.

\begin{figure*}[h!]
\centering
\includegraphics[width=\linewidth]{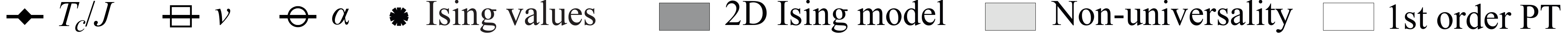}
\subfloat[]{\includegraphics[width=0.49\linewidth]{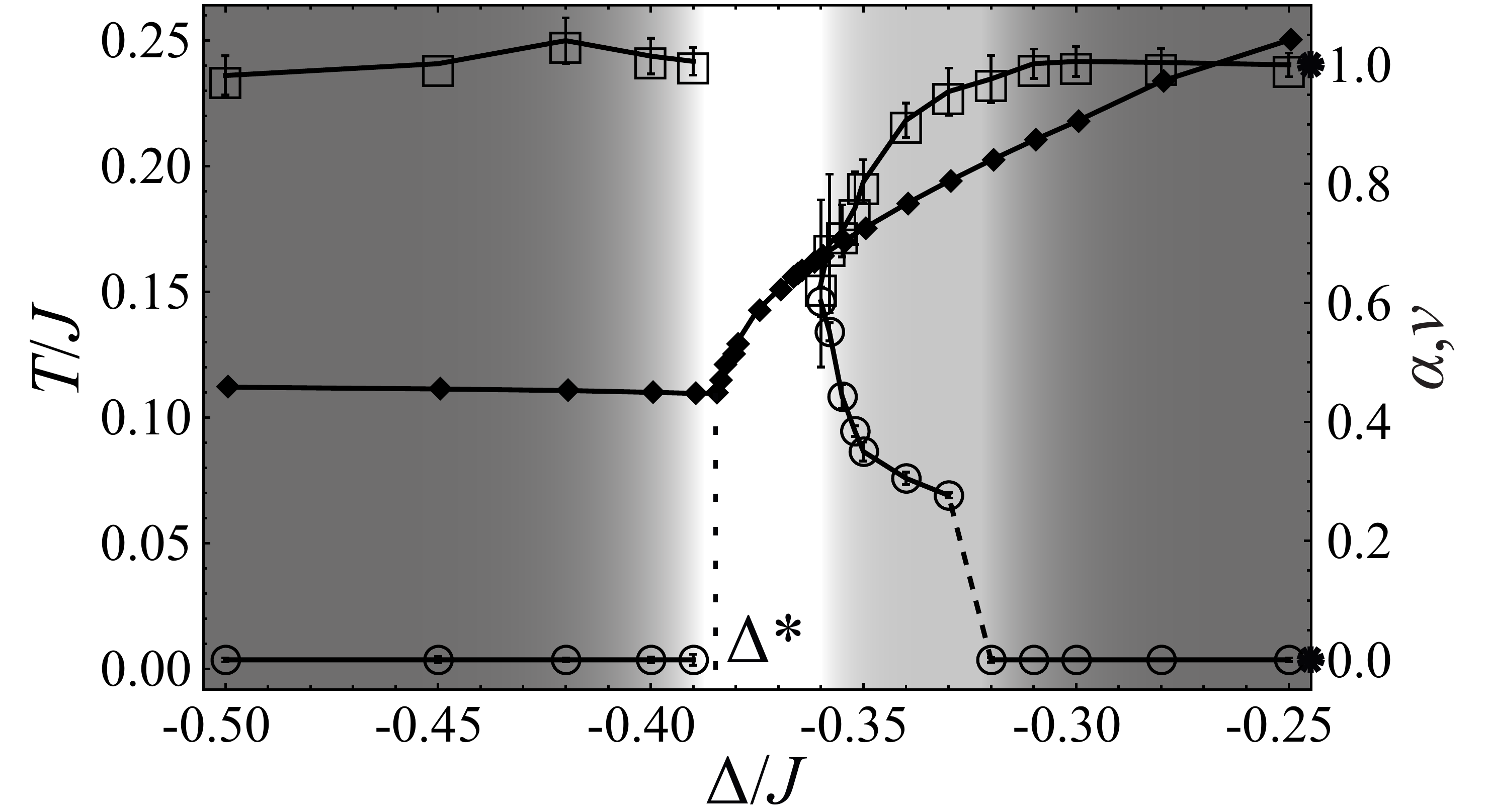}
\label{V02}}
\hfil
\subfloat[]{\includegraphics[width=0.49\linewidth]{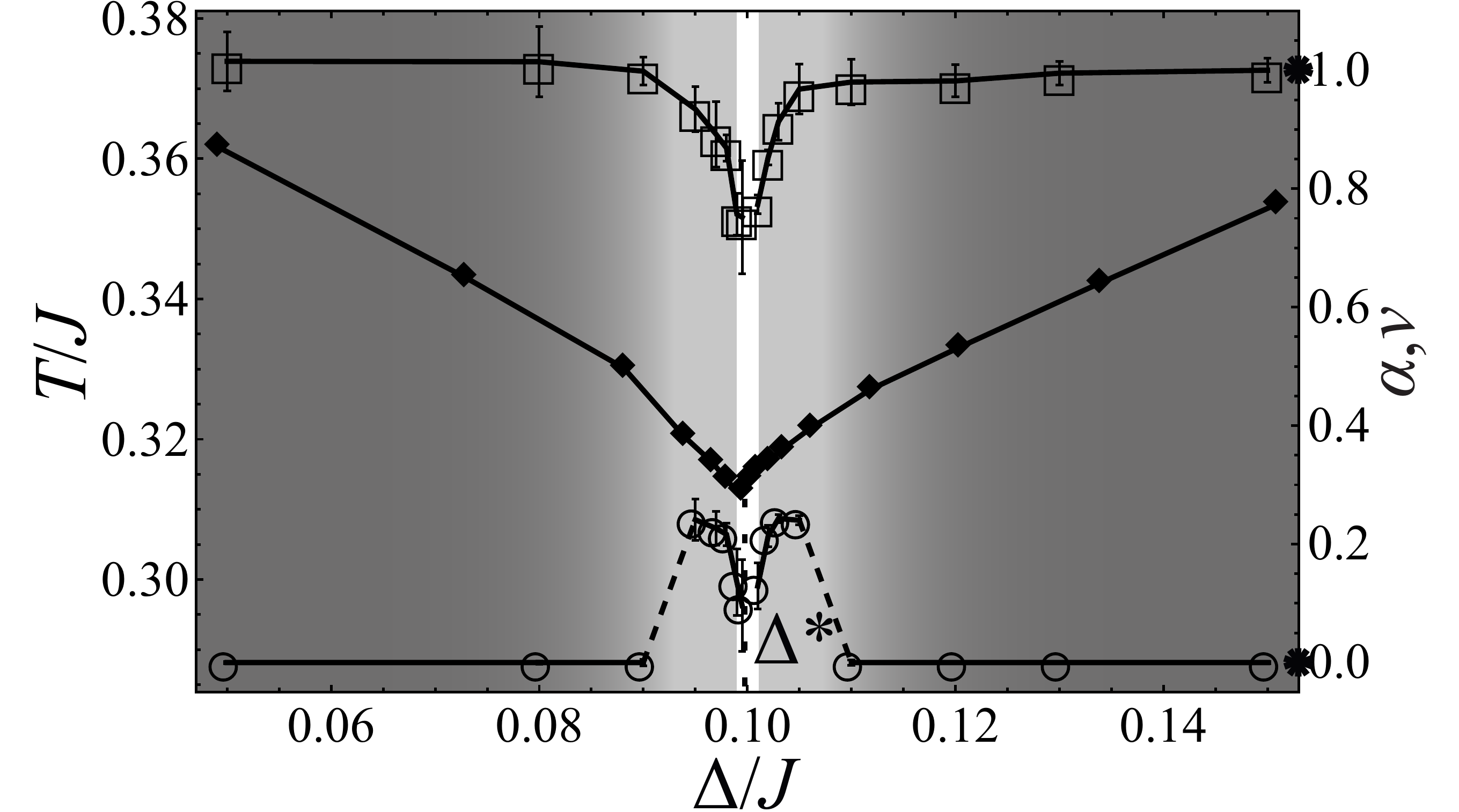}
\label{V12}}
\caption{Dependencies of critical temperature $T_c/J$, critical exponents $\alpha$ and $\nu$ on ``single-ion anisotropy'' parameter $\Delta/J$ for (a) $V/\tilde{J}=0.2$, $n=0$; (b) $V/\tilde{J}=1.2$, $n=0$}

\includegraphics[width=\linewidth]{FIG3.pdf}
\subfloat[]{\includegraphics[width=0.49\linewidth]{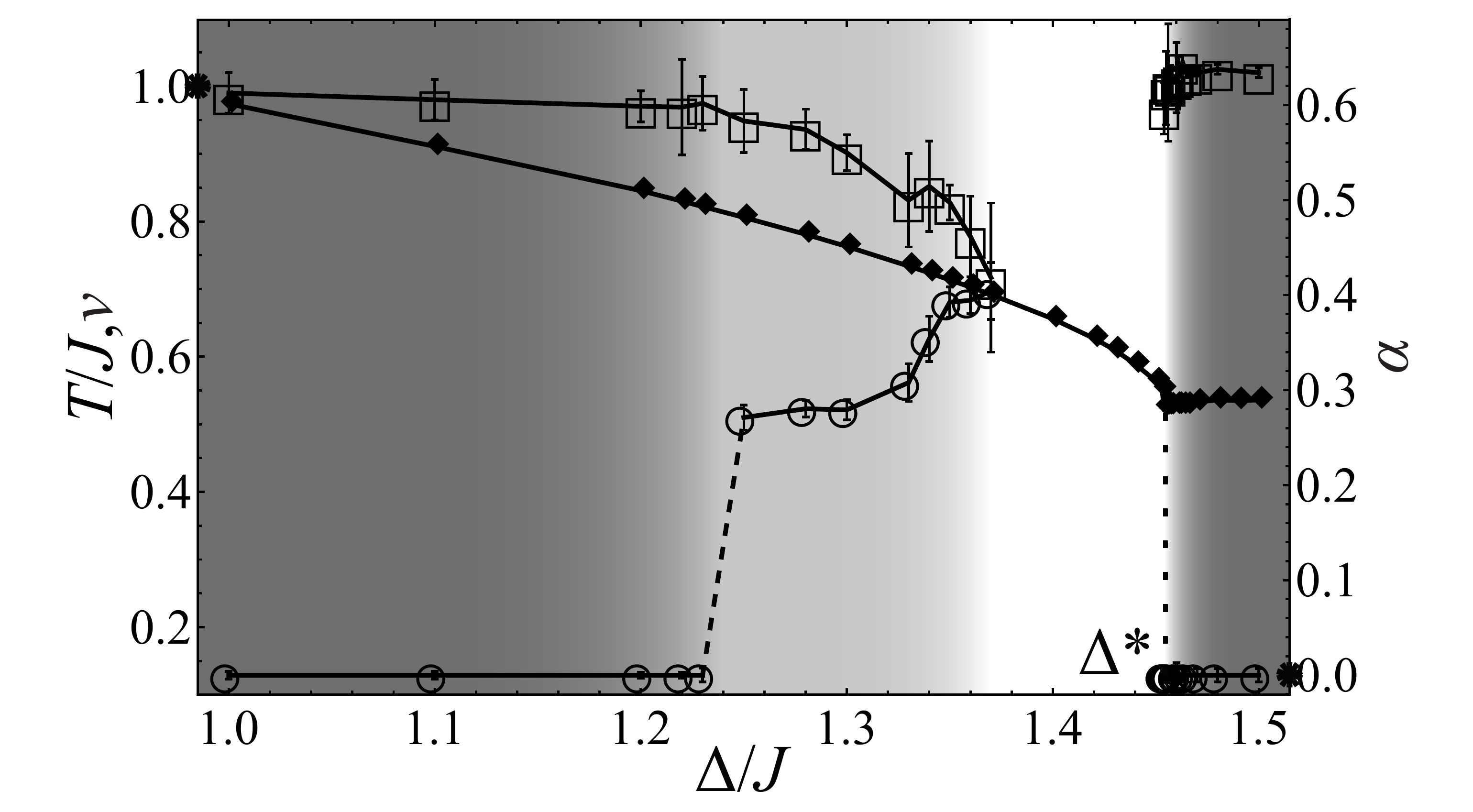}
\label{V4}}
\hfil
\subfloat[]{\includegraphics[width=0.49\linewidth]{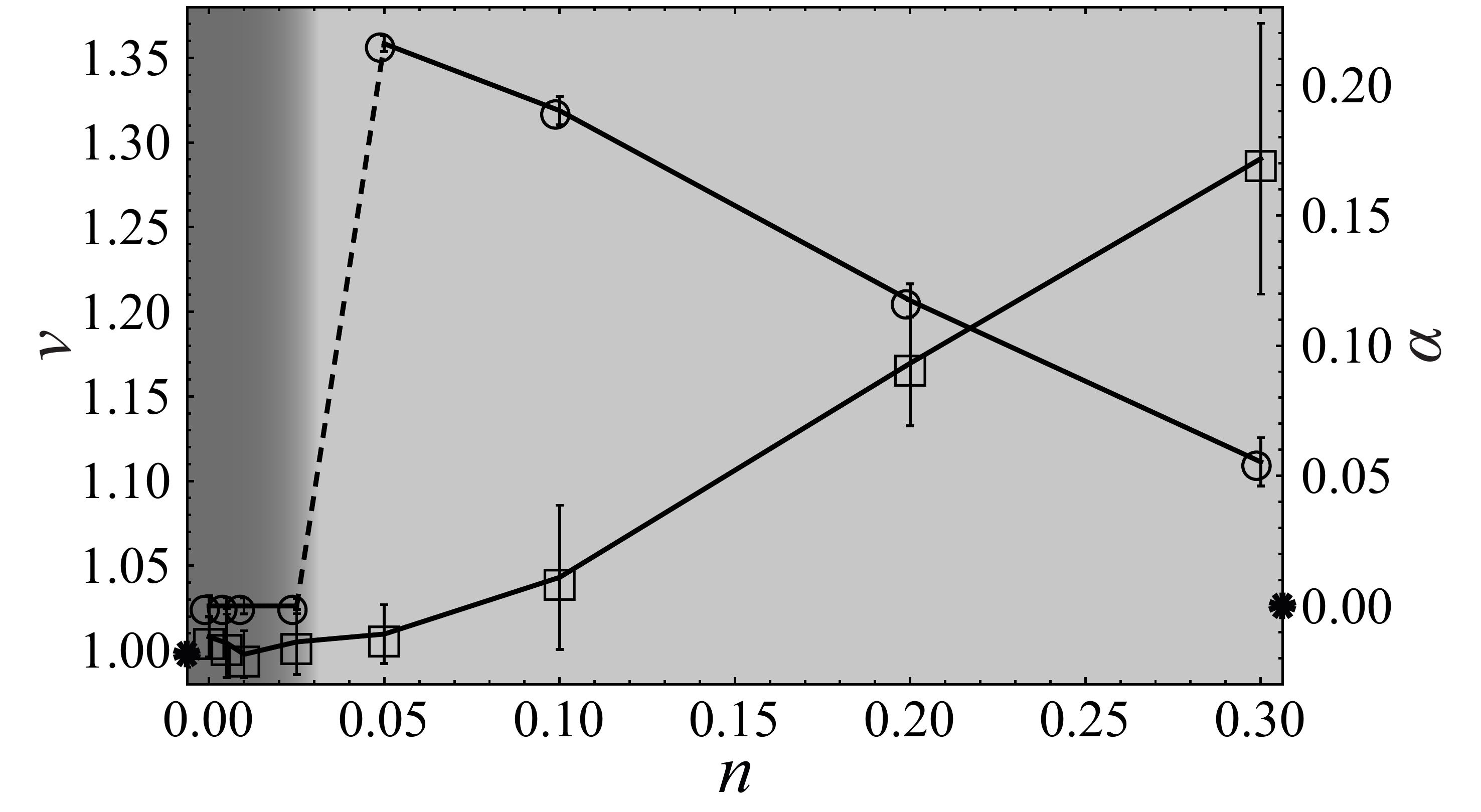}
\label{V04}}
\caption{Dependencies of critical temperature $T_c/J$, critical exponents $\alpha$ and $\nu$ on (a) ``single-ion anisotropy'' parameter $\Delta/J$ for $V/\tilde{J}=4.0$, $n=0$; (b) charge density $n$ for $V/\tilde{J}=0.4$, $\Delta/J=0$ ($T_c/J$ is not shown)}
\end{figure*}

To explore the effects of frustration and non-magnetic impurities on the critical properties of our system, we obtained the dependencies of critical temperatures $T_c$, critical exponents of the specific heat $\alpha$ and correlation length $\nu$ on ``single-ion anisotropy'' $\Delta/J$ and charge density $n$ for different relationships between parameters of pseudospin and spin interactions $V/\tilde{J}$. We present these dependencies on $\Delta/J$ for $V/\tilde{J}=0.2$, $n=0$ in Fig~\ref{V02}. The frustration point is denoted by $\Delta^*$.
Ground-state configurations are given by antiferromagnetic (AFM) order for $\Delta/J>\Delta^*$ and charge order (CO) for $\Delta/J<\Delta^*$. It can be seen that exponents take the values of 2D Ising universality class ($\nu_{Ising}=1$, $\alpha_{Ising}=0$, marked with black stars) for $\vert \Delta/J \vert \gg \Delta^*$.
The area where the phase transition belongs to the 2D Ising universality class is marked by dark gray. The exponents vary monotonically as $\Delta/J$ decreases to $\Delta^*$, i.e. the critical behavior near the frustration point $\Delta^*$ becomes non-universal. This area is marked by light gray. A first-order phase transition occurs in the area marked by white, and the critical exponents are not calculated.

Figures~\ref{V12}--\ref{V04} are designed in a similar way. They show the dependencies of the critical temperature $T_c$ and critical exponents $\alpha$ and $\nu$ on $\Delta/J $ for $n = 0$, $V/\tilde{J} = 1.2$ (Fig.~\ref{V12}) and $ V/\tilde{J} = 4 $ (Fig.~\ref {V4}). The presented data allow us to conclude again that the model belongs to the 2D Ising universality class far from the frustration point, and nonuniversal critical behavior and first-order PT are observed near the frustration point. The difference appears when approaching the boundary between strong and weak spin exchange $V/\tilde{J} = 1$. Firstly, the region with first-order PT decreases, and secondly, the exponent of the specific heat $\alpha$ begins to decrease rather than increase.
For $V/\tilde{J}<1 $ non-universal critical behavior and first-order PT appeared for the transition to the AFM state. For $V/\tilde{J}>1$ this occurs for the transition to the CO state. Since the area of non-universality is adjacent to the area with first-order PT, there is a possibility that evidence of the first or second-order PT occurs only in a certain scale of the system. For example, the instability of a first-order signature in energy histograms was found in~\cite{Kalz}, which examines the Ising model with competing interactions and also shows non-universal critical behavior and area with first-order phase transitions. Therefore, frustrated systems require extremely careful and accurate analysis of critical behavior.

In addition, we studied the influence of charged impurities on the critical behavior for various $V/\tilde{J}$. Fig.~\ref{V04} shows the dependence of $\alpha$ and $\nu$ ($T_c/J$ is not shown) on density of charged impurities $n$ for $V/\tilde{J} = 0.4$, $\Delta/J = 0$.
For small $n$, the exponents take the Ising values. However, as the density increases, the critical exponents begin to depend on $n$, which also indicates the non-universal critical behavior. This result coincides with those for other ratios $V/\tilde{J}$ and for the transition to the CO state. A possible explanation is that the presence of interacting impurities leads to the dissipation of critical fluctuations responsible for phase transitions. This phenomenon is consistent, for example, with Dotsenko's results for the 2D Ising model with random bonds~\cite{Dotsenko}. It is important to note that non-magnetic impurities do not change the order of PT.

\section{Conclusion}

We explored the critical properties of the spin-pseudospin model, which describes a 2D Ising magnet with non-magnetic impurities and competing charge and spin interactions. We performed
extensive Monte-Carlo simulations to calculate the critical exponents of the specific heat $\alpha$ and correlation length $\nu$ in the framework of the finite-size scaling theory. The calculations were performed for a wide range of model parameters, such as ``single-ion anisotropy'' $\Delta/J$ (local density-density correlation parameter), density of nonmagnetic impurities $n$, and the relationship between between pseudospin and spin exchange interactions $V/\tilde{J}$. Energy histogram method enabled us to determine the order of phase transitions.

The 2D Ising universality class is revealed to remain far from the frustration point and at low densities of non-magnetic impurities. We found that frustration and non-magnetic impurities lead to non-universal critical behavior. Moreover, first-order phase transitions are observed near the frustration point. The features of the non-universal critical behavior, as well as the width of the area with first-order phase transitions, depend on the relationship between the spin and pseudospin interactions.

\section*{Acknowledgment}

This work was supported in part by Competitiveness Enhancement Program of the Ural Federal University (Act 211 of the Government of the Russian Federation, Agreement N. 02.A03.21.0006 and CEP 3.1.1.1-20), and also the Ministry of Education and Science of the Russian Federation (project FEUZ-2020-0054). We also would like to thank Alexander Moskvin for fruitful discussions.

\end{document}